# Propagation of defects in doped magnetic materials of different dimensionality

A. Furrer[1], A. Podlesnyak[2], K.W. Krämer[3], J.P. Embs[1], V. Pomjakushin[1], and Th. Strässle[1]

[1]Laboratory for Neutron Scattering, Paul Scherrer Institut, CH-5232 Villigen PSI, Switzerland

[2]Quantum Condensed Matter Division, Oak Ridge National Laboratory, Oak Ridge, TN 37831-6473, USA

[3]Department of Chemistry and Biochemistry, University of Bern, CH-3012 Bern, Switzerland

Defects intentionally introduced into magnetic materials often have a profound effect on the physical properties. Specifically tailored neutron spectroscopic experiments can provide detailed information on both the local exchange interactions and the local distances between the magnetic atoms around the defects. This is demonstrated for manganese dimer excitations observed for the magnetically diluted three- and two-dimensional compounds $KMn_xZn_{1-x}F_3$ and $K_2Mn_xZn_{1-x}F_4$, respectively. The resulting local exchange interactions deviate up to 10% from the average, and the local Mn-Mn distances are found to vary stepwise with increasing internal pressure due to the Mn/Zn substitution. Our analysis qualitatively supports the theoretically predicted decay of atomic displacements according to $1/r^2$, $1/r$, and constant (for three-, two-, and one-dimensional compounds, respectively) where $r$ denotes the distance of the displaced atoms from the defect.

## I. INTRODUCTION

Atoms in perfect crystals are arranged in a periodic lattice. However, perfect crystals do not exist in nature. All crystals have some inherent defects, *e.g.*, vacancies or substitutional atoms due to the limited purity of the material. Moreover, defects are often intentionally introduced into the crystal in order to manipulate its physical properties. Doping has become an important tool for magnetic materials in the context of high-temperature superconductivity [1] and quantum magnetism [2] as well as for semiconductors to induce ferromagnetism [3]. Defects produce local lattice distortions, *i.e.,* the atoms around the defect are displaced from their crystallographic equilibrium positions. As a consequence, the exchange interactions around the defects deviate from the average exchange. Experimental information on local exchange interactions and local structures is important to understand the physical mechanisms of the defect crystals. Some commonly used techniques for local structure determination are listed in Ref. 4. From the theoretical side, the defect problem was investigated by Krivoglaz [5] who predicted that the atomic displacements in three-dimensional crystals decay asymptotically as *1/$r^2$* (*r* denotes the distance of the displaced atoms from the defect). For defects in two-dimensional crystals the atomic displacements are expected to decrease according to *1/r*, and in one-dimensional crystals the atomic displacements do not decrease with increasing distance *r* at all. The latter was recently verified by an inelastic neutron scattering (INS) study of $Mn^{2+}$ dimer excitations in the one-dimensional paramagnetic compound $CsMn_xMg_{1-x}Br_3$ [4], whereas to our knowledge no experimental proofs have been given so far for the three- and two-dimensional cases.

The present work provides experimental evidence for the theoretically predicted *1/$r^2$* and *1/r* laws for $KMn_xZn_{1-x}F_3$ and $K_2Mn_xZn_{1-x}F_4$ (x=0.10), respectively, which are three- and two-dimensional magnetic compounds as confirmed by INS experiments of the spin-wave dispersion in KMnF$_3$ [6] and K$_2$MnF$_4$ [7]. The different magnetic dimensionality is also reflected in the elastic properties. Both compounds are ionic crystals, thus the elastic energy is mainly governed by Coulomb interactions (apart from the repulsive potential) depending on the Madelung constant which is a property of the crystal structure. KMnF$_3$ is built up from corner-sharing MnF$_6$ octahedra along all three crystallographic symmetry directions. This is also true for K$_2$MnF$_4$ along the x and y directions, but along the tetragonal z direction there is an additional K-F layer inserted. Accordingly, the elastic energy along the z direction is considerably weaker than in the x-y plane, since K is monovalent in contrast to the divalent Mn and the Mn-F distances are significantly shorter than the K-F distances. In fact, calculations based on Coulomb interactions show that the lattice energy associated with the K-F layers is typically four times smaller than that of the Mn-F layers, thus it is justified to describe the elastic properties of K$_2$MnF$_4$ by a quasi-two-dimensional model.

We use the INS technique to probe the energy *E* of the singlet-triplet splitting of Mn$^{2+}$ dimers as described in Ref. 4. For KMn$_x$Zn$_{1-x}$F$_3$ earlier INS experiments performed with x=0.18 [8] and x=0.10 [9] showed that *E*=0.82 meV at T=10 K, and *E* was found to slightly decrease with increasing temperature [9]. On the other hand, for K$_2$Mn$_x$Zn$_{1-x}$F$_4$ corresponding data are not available so far, but *E*≈0.8 meV is expected from Ref. 7. Our analysis concerning local structural and magnetic effects is based on the linear law *dJ/dR=α* (valid for *dR<<R*), where *J=E/2* is the exchange coupling of magnetic atoms and *R* their interatomic distance. The quantity *dJ/dR* can be derived from the temperature dependence of *J* and *R* as described in Ref. 10, thus we performed corresponding neutron scattering experiments for both compounds.

The highly resolved INS spectra associated with the singlet-triplet splitting of Mn$^{2+}$ dimers in KMn$_{0.10}$Zn$_{0.90}$F$_3$ and K$_2$Mn$_{0.10}$Zn$_{0.90}$F$_4$ exhibit remarkable fine structures, which we attribute to local structural and magnetic effects. More specifically, we find that the local exchange interactions can deviate up to 10% from the average. In addition, we derive the local intradimer distances which are found to decrease stepwise with increasing internal pressure (resulting from the Mn/Zn substitution and therefore also referred to as chemical pressure).

**II. SYNTHESIS AND CHARACTERIZATION OF SAMPLES**

Samples of $KMn_{0.10}Zn_{0.90}F_3$ and $K_2Mn_{0.10}Zn_{0.90}F_4$ were synthesized from KF, $MnF_2$, and $ZnF_2$. KF (Alfa Aesar, 99.99%) was dried first at 200°C in an Ar/HF gas stream and then in high vacuum. $ZnF_2$ was prepared from ZnO (Alfa, 99.995%). The oxide was dissolved in concentrated $HNO_3$ (Merck, p.a.), the solution dried, and the fluoride prepared by repeated adding of concentrated HF acid (Merck, suprapur) and drying three times. Afterwards the powder was heated to 450°C in an Ar/HF gas stream for complete fluorination. $MnF_2$ was synthesized from $MnCO_3$ (Alfa, 99.985%) by repeated adding of concentrated HF acid (Merck, suprapur) and drying for three times. Finally, the powder was heated to 450°C in an Ar/HF gas stream. In a dry box (MBraun, Munich) the starting materials were filled into platinum ampoules which then were sealed off by arc welding under 300 mbar He. Batches had a total weight of typically 15 g. For $KMn_{0.10}Zn_{0.90}F_3$ a stoichiometric mixture of the starting materials was used. Since $K_2ZnF_4$ melts incongruently, a molar ratio of 0.73 KF, 0.027 $MnF_2$, and 0.243 $ZnF_2$ was used for the $K_2Mn_{0.10}Zn_{0.90}F_4$ sample. The Pt ampoules were placed in a tube furnace under Ar gas and heated to 920°C and 800°C for $KMn_{0.10}Zn_{0.90}F_3$ and $K_2Mn_{0.10}Zn_{0.90}F_4$, respectively. The temperature was kept for 10 hours followed by slow cooling with 5 K/h to room temperature. The composition and phase purity was checked by powder X-ray diffraction. $KMn_{0.10}Zn_{0.90}F_3$ crystallizes in the cubic space group *Pm3m* [11] and turned out to be phase pure. $K_2Mn_{0.10}Zn_{0.90}F_4$ crystallizes in the tetragonal space group

*I4/mmm* [12], but the diffraction pattern showed some extra lines associated with KF excess.

Neutron diffraction and spectroscopy experiments were carried out at the spallation neutron source SINQ at PSI Villigen with use of the instruments HRPT and FOCUS, respectively, in order to determine the temperature dependence of the parameters $R$ and $J$. Figure 1 shows energy spectra of neutrons scattered from $K_2Mn_{0.10}Zn_{0.90}F_4$, which are very similar to those obtained for $KMn_{0.10}Zn_{0.90}F_3$ [9]. The well defined inelastic peaks can easily be attributed to particular $Mn^{2+}$ dimer (D) and trimer (T) transitions with energies given by the subscript zJ. The unresolved shoulder around 0.4 meV is due to the lowest $Mn^{2+}$ tetramer transition. The particular multimer nature of all these transitions was confirmed by the characteristic T and Q dependence of the intensities [13], where Q is the modulus of the scattering vector **Q**. The transition $D_{2J}$ corresponds to the singlet-triplet splitting which will be further investigated in the present work. All the results are listed in Tables I and II, from which we derive $dJ/dR=\alpha=3.3(6)$ meV/Å for $KMn_{0.10}Zn_{0.90}F_3$ (using the T=20 K and T=200 K data) and $\alpha=2.6(4)$ meV/Å for $K_2Mn_{0.10}Zn_{0.90}F_4$ (using the T=10 K and T=100 K data).

## III. HIGH-RESOLUTION INS EXPERIMENTS

High-resolution INS experiments were performed with use of the time-of-flight spectrometer CNCS [14] at the spallation neutron source SNS at Oak Ridge National Laboratory. The samples were enclosed in Al cylinders (12 mm diameter, 45 mm height) and placed into a He cryostat to achieve temperatures T>2 K. The energy of the incoming neutrons was 1.55 meV, giving rise to an unprecedented energy resolution of Gaussian shape with FWHM=11 μeV at $E\approx0.8$ meV. Data were collected for scattering vectors **Q** with moduli $0.5\leq Q\leq1.3$ Å$^{-1}$.

## A. KMn$_{0.10}$Zn$_{0.90}$F$_3$

The partial substitution of Zn$^{2+}$ ions by 10% Mn$^{2+}$ ions results in the statistical creation of Mn$^{2+}$ monomers (53.1%), dimers (20.9%), trimers (8.2%), tetramers (2.8%), etc. Ground-state transitions associated with all these multimers were clearly observed in earlier INS experiments [9]. For the present application we focus on the lowest-lying Mn$^{2+}$ dimer excitation corresponding to the singlet-triplet transition. An energy spectrum observed at T=2 K and shown in Fig. 2 exhibits a remarkable fine structure which could not be observed in earlier INS experiments [8,9]. A visual inspection of the data suggests that the spectrum corresponds to a superposition of up to seven individual lines of Gaussian shape with almost equidistant energy spacings. We analyzed the data in this manner by a least-squares fitting procedure. The adjustable parameters were a linear background, the positions and the amplitudes of the individual lines as well as an overall linewidth which was kept equal for all lines. The initial positions were set at equidistant energies from 0.73, 0.76, ..., 0.91 meV with initial amplitudes given by the observed intensities at these positions. The least-squares fitting procedure rapidly converged to the results displayed as dashed lines in Fig. 2.

What is the origin of the presence of seven individual lines? Obvious mechanisms are the single-ion or exchange anisotropies as well as next-nearest-neighbor exchange couplings, but all these interactions split the singlet-triplet excitation into three lines at most and therefore cannot explain our findings. Moreover, the next-nearest-neighbor exchange parameter was reported to be smaller than 1-2 µeV [9] and therefore cannot produce an energy distribution with a total width of about 200 µeV (see Fig. 2). Nevertheless, these mechanisms are certainly present and manifest themselves in the overall linewidth with FWHM=27(3) µeV resulting from the least-squares fitting procedure (thereby considerably exceeding the instrumental energy resolution of 11 µeV).

A possible mechanism for the observed fine structure is due to local structural inhomogeneities. More specifically, the statistical substitution of $Zn^{2+}$ by $Mn^{2+}$ ions exerts some internal pressure, since the ionic radii of the $Mn^{2+}$ (high spin) and $Zn^{2+}$ ions are different with $r_{Mn}$=0.83 Å > $r_{Zn}$=0.74 Å [15], so that the atomic positions have to rearrange in the vicinity of the dopant $Mn^{2+}$ ions. We explain this picture with the help of the local structure of an isolated $Mn^{2+}$ dimer as sketched in Fig. 3. There are n=26 nearest-neighbor positions around the dimer which can be occupied either by $Zn^{2+}$ or $Mn^{2+}$ ions. The probability for having m $Mn^{2+}$ ions at the nearest-neighbor positions is given by

$$p_m(x) = \binom{n}{m} x^m (1-x)^{n-m}. \qquad (1)$$

The probabilities $p_m(x)$ calculated for the $Mn^{2+}$ concentration x=0.10 first increase with increasing m, have a maximum for m=2, and then decrease until $p_m(x)$<1% for m≥7. The probabilities $p_m(x)$ are indicated for 0≤m≤6 as vertical bars in Fig. 2. They are in rather good agreement with the amplitudes of the individual lines with a goodness of fit $\chi^2$=1.9 [16], thus there is no need to extend the local structural model beyond the nearest-neighbor positions. Actually this conclusion is in line with the theoretically predicted *1/r²* law for atomic displacements around a defect in three-dimensional crystals [5].

**B. $K_2Mn_{0.10}Zn_{0.90}F_4$**

According to statistics, the $Mn^{2+}$(10%)/$Zn^{2+}$(90%) substitution gives rise to $Mn^{2+}$ monomers (65.6%), dimers (21.3%), trimers (6.4%), tetramers (1.6%), etc. We again focus on the lowest-lying $Mn^{2+}$ dimer excitation. Fig. 4 shows the corresponding energy spectrum taken with the same experimental parameters as used for Fig. 2. The data clearly exhibit some fine structure, although less pronounced than those of Fig. 2. We treat the data in terms of the local structure around an isolated $Mn^{2+}$ dimer as visualized in Fig. 5. There are n=10 nearest-neighbor positions around the dimer. According to Eq. (1), the probabilities $p_m(x)$ are smaller than 1% for m≥5, thus we analyzed the data by a superposition of five individual lines of Gaussian shape, similar to the least-squares fitting procedure described in Sec. III.A. The results are shown in Fig. 4(a). The probabilities $p_m(x)$ for 0≤m≤4 are indicated by vertical bars, however, there is no agreement with the amplitudes of the individual lines (goodness of fit $\chi^2$=36). We therefore extend our analysis to include in addition the 14 next-nearest-neighbors around the isolated $Mn^{2+}$ dimer (see Fig. 5). In this case we have n=24 in Eq. (1), and probabilities $p_m(x)$ up to m=6 have to be considered. The subdivision of the spectrum into seven individual lines resulting from the least-squares analysis is shown in Fig. 4(b). Now the agreement between the amplitudes of the individual lines and the corresponding probabilities $p_m(x)$ for 0≤m≤6 indicated by vertical bars is drastically improved, with a goodness of fit $\chi^2$=2.6. The extension of the local structural model beyond the nearest-neighbors is not surprising, since the atomic displacements around a defect in two-dimensional crystals are governed by a *1/r* law according to theory [5].

**IV. DISCUSSION OF RESULTS**

All the individual lines shown in Figs 2 and 4(b) labelled by m=0,1,2,...,6 correspond to specific $Mn^{2+}$ dimer configurations, with particular exchange coupling parameters $J_m$ being proportional to the corresponding energies $E_m$ of the lines according to the Heisenberg Hamiltonian

$$H_m = -2J_m \mathbf{s_1} \cdot \mathbf{s_2} , \qquad (2)$$

where $s_i$ denotes the spin operator of the $Mn^{2+}$ ions. The energies $E_m$ are increasing from m=0 to m=6, thus the parameters $J_m=-E_m/2$ are increasing as well. Based on the linear law $dJ/dR$ determined in Sec. II we can now calculate the shrinking of the local intradimer Mn-Mn distance $R$ as a function of internal pressure induced by the Mn/Zn substitution. The lattice parameters $a$ were taken as the mean local distances.

Table III lists the resulting values of $J_m$ and $R_m$. The local exchange interactions $J_m$ are found to deviate up to 10% from the average exchange. The local distances $R_m$ are distributed at almost equidistant steps $\Delta R=R_m-R_{m-1}=-0.0044(7)$ Å for the three-dimensional compound $KMn_{0.10}Zn_{0.90}F_3$ and $\Delta R=-0.0031(6)$ Å for the quasi-two-dimensional compound $K_2Mn_{0.10}Zn_{0.90}F_4$. These values are roughly equal within experimental error as expected, since the structures of both compounds are characterized by corner-sharing $MF_6$ (M=Mn or Zn) octahedra. The size of the lattice distortions $\Delta R$ can be related to the elastic stiffness tensor, for which the dominant element is $c_{11}=132$ GPa for perovskite structures [17]. For the one-dimensional compound $CsMn_xMg_{1-x}Br_3$ the shrinking was reported to be $\Delta R=-0.0022(4)$ Å [4]. Here the hexagonal structure is characterized by face-sharing $MBr_6$ (M=Mn or Mg) octahedra along the $c$ axis. The dominant element of the elastic stiffness tensor is $c_{33}=64$ GPa [10], which most likely explains the considerably smaller size of $\Delta R$.

## V. CONCLUDING REMARKS

Our analysis of the fine structure of $Mn^{2+}$ dimer excitations qualitatively supports the theoretically predicted $1/r^2$ and $1/r$ decay laws of local atomic displacements resulting from substitutional defects in three- and two-dimensional crystals, respectively. The agreement between the experimental and the calculated data is remarkably good, in spite of the rather simple statistical model which is governed by the number *m* of $Mn^{2+}$ ions and not by their specific arrangement around the isolated dimer. Obviously the substitution of $Zn^{2+}$ ions by $Mn^{2+}$ ions creates a local internal pressure whose nature is essentially hydrostatic and which is most likely mediated by the elastic stiffness tensor.

Currently, investigations of quantum spin systems are of increasing interest in the search for exotic ground states. A very informative approach has been to study the modifications of the physical properties induced by the controlled introduction of defects in the materials, whose precise behavior depends on the value of the spin, the anisotropy, the dimensionality of the material, and the strengths and signs of the magnetic couplings [18,19]. The present work adds important information on the two latter quantities. We have shown that the exchange interaction in doped materials is no longer uniformly distributed, but exhibits marked differences around the defects. Furthermore, the spatial extension of the „distorted" region around the defects strongly increases with lowering the dimensionality of the spin system. In particular, these effects cannot be neglected in studies of doped spin-chain materials like $SrCuO_2$ [20] and $SrCu_2O_3$ [21] with an exchange interaction of the order of J≈200 meV. Even if the magnetic couplings around the defects are modified only by a few percent, these differences match the relevant energy scale at low temperatures and therefore have to be considered in the data analysis.

**ACKNOWLEDGMENT**

The assistance of D. Biner (University of Bern) in the synthesis of the samples is gratefully acknowledged. Part of this work was performed at the Swiss Spallation Neutron Source (SINQ), Paul Scherrer Institut (PSI), Villigen, Switzerland. Research at Oak Ridge National Laboratory's Spallation Neutron

Source was supported by the Scientific User Facilities Division, Office of Badic Energy Sciences, US Department of Energy.

TABLE I. Temperature dependence of the lattice parameter *a* (present work) and the dimer excitation energies $E_{zJ}$ (taken from Ref. 9) for $KMn_{0.10}Zn_{0.90}F_3$. The diffraction experiments were performed with a neutron wavelength λ=1.494 Å.

| T [K] | a [Å] | $E_{2J}$ [meV] | $E_{4J}$ [meV] | $E_{6J}$ [meV] | $E_{8J}$ [meV] |
|---|---|---|---|---|---|
| 10 | 4.05733(2) | | | | |
| 20 | | 0.817(2) | 1.604(3) | 2.408(4) | 3.198(13) |
| 50 | 4.05752(2) | 0.812(2) | 1.596(2) | 2.396(3) | 3.171(6) |
| 100 | 4.05854(2) | 0.802(3) | 1.582(2) | 2.379(3) | 3.148(4) |
| 150 | 4.06051(2) | 0.796(3) | 1.562(3) | 2.338(5) | 3.094(5) |
| 200 | 4.06314(2) | 0.779(6) | 1.538(4) | 2.303(4) | 3.039(8) |

TABLE II. Temperature dependence of the lattice parameters $a$ and $c$ and the dimer excitation energies $E_{zJ}$ for $K_2Mn_{0.10}Zn_{0.90}F_4$ determined in the present work. The diffraction experiments were performed with a neutron wavelength $\lambda=1.494$ Å.

| T [K] | $a$ [Å] | $c$ [Å] | $E_{2J}$ [meV] | $E_{4J}$ [meV] | $E_{6J}$ [meV] |
|---|---|---|---|---|---|
| 1.6 | | | 0.828(1) | | |
| 5 | 4.04850(5) | 13.07255(23) | | | |
| 10 | | | 0.828(1) | 1.648(1) | 2.462(4) |
| 50 | | | 0.823(2) | 1.640(2) | 2.458(2) |
| 100 | 4.05168(6) | 13.07716(26) | 0.814(3) | 1.616(2) | 2.417(3) |
| 150 | 4.05492(6) | 13.08587(26) | | | |
| 200 | 4.05869(7) | 13.09734(28) | | | |
| 250 | 4.06263(7) | 13.10990(30) | | | |
| 290 | 4.06314(2) | 13.12018(31) | | | |

TABLE III. Analysis of experimental data observed for $KMn_{0.10}Zn_{0.90}F_3$ and $K_2Mn_{0.10}Zn_{0.90}F_4$. The exchange couplings $J_m$ and the local Mn-Mn distances $R_m$ associated with each line m were derived as explained in the text. For both $J_m$ and $R_m$ relative error bars are given.

|   | $KMn_{0.10}Zn_{0.90}F_3$ | | $K_2Mn_{0.10}Zn_{0.90}F_4$ | |
|---|---|---|---|---|
| m | $J_m$ [meV] | $R_m$ [Å] | $J_m$ [meV] | $R_m$ [Å] |
| 0 | -0.364(2) | 4.0697(8) | -0.390(3) | 4.0566(8) |
| 1 | -0.382(2) | 4.0643(7) | -0.398(2) | 4.0535(6) |
| 2 | -0.396(1) | 4.0591(6) | -0.406(1) | 4.0504(4) |
| 3 | -0.411(1) | 4.0546(5) | -0.414(1) | 4.0473(4) |
| 4 | -0.425(1) | 4.0512(6) | -0.423(1) | 4.0439(5) |
| 5 | -0.440(2) | 4.0467(7) | -0.430(2) | 4.0412(6) |
| 6 | -0.452(3) | 4.0431(8) | -0.438(3) | 4.0381(7) |

**Figure Captions**

FIG. 1. Energy spectra of neutrons scattered from $K_2Mn_{0.10}Zn_{0.90}F_4$. The incoming neutron energy was 4.2 meV. The arrows mark the multimer transitions (D=dimer, T=trimer) with energies indicated by the subscript zJ.

FIG. 2. (Color online) Energy distribution of the $Mn^{2+}$ singlet-triplet dimer transition observed for $KMn_{0.10}Zn_{0.90}F_3$ at T=2 K. The lines are the result of a least-squares fitting procedure as explained in the text. The vertical bars correspond to the probabilities $p_m(x)$ predicted by Eq. (1) with n=26.

FIG. 3. (Color online) Three-dimensional sketch of the structure of $KMn_xZn_{1-x}F_3$ (only the Mn and Zn positions are shown). The triangles denote the 26 nearest-neighbor positions around the central $Mn^{2+}$ dimer marked by spheres.

FIG. 4. (Color online) Energy distribution of the $Mn^{2+}$ singlet-triplet dimer transition observed for $K_2Mn_{0.10}Zn_{0.90}F_4$ at T=2 K. The lines are the result of a least-squares fitting procedure as explained in the text. The vertical bars correspond to the probabilities $p_m(x)$ predicted by Eq. (1). The goodness of fit $\chi^2$ is a measure of the agreement between the probabilities $p_m(x)$ and the amplitudes of the individual lines. (a) Statistical model with n=10 nearest-neighbor positions. (b) Statistical model with n=24 nearest-neighbor and next-nearest-neighbor positions.

FIG. 5. (Color online) Two-dimensional sketch of the structure of $K_2Mn_xZn_{1-x}F_4$ (only the Mn and Zn positions are shown). The full and open triangles denote the 10 nearest-neighbor and 14 next-nearest-neighbor positions, respectively, around the central $Mn^{2+}$ dimer marked by spheres.

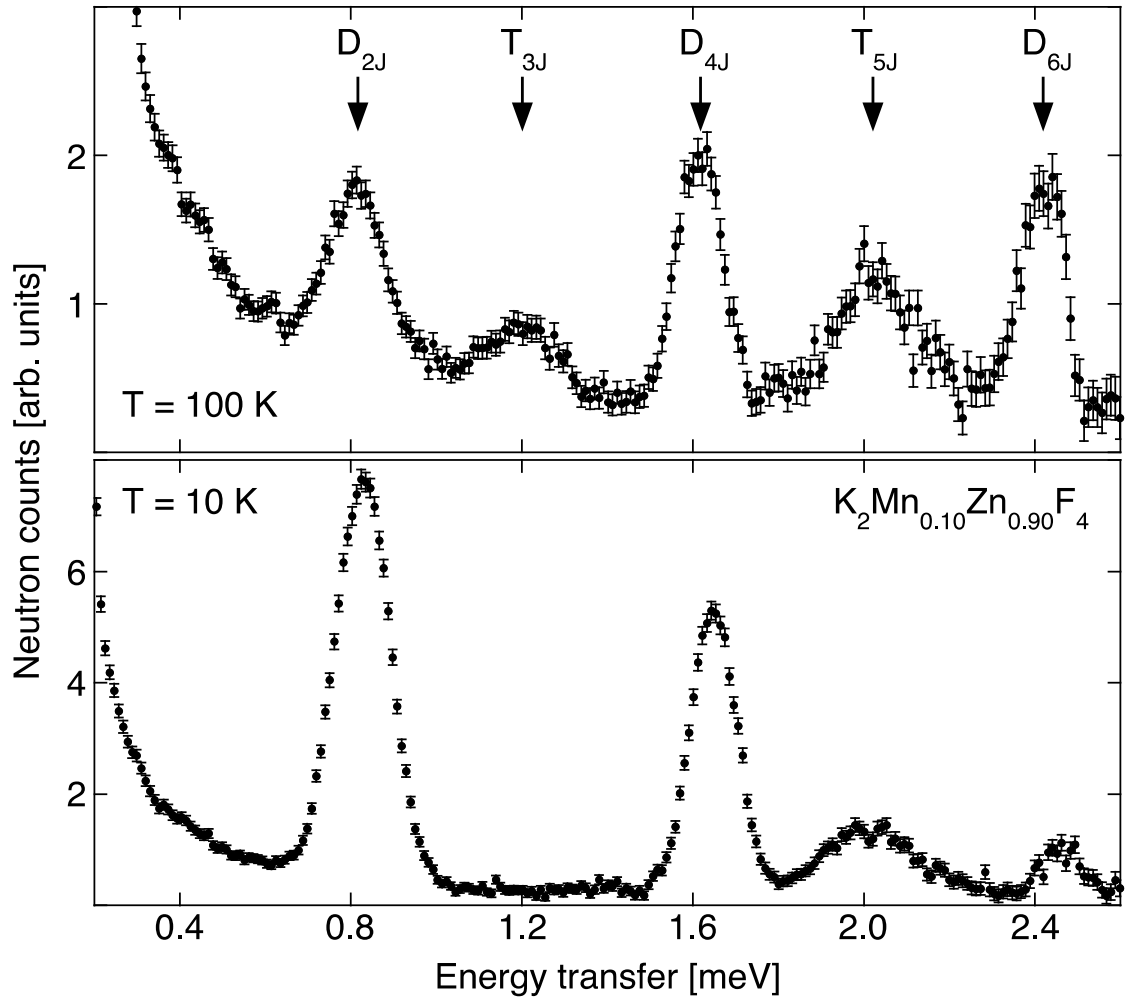

FIG. 1. Energy spectra of neutrons scattered from $K_2Mn_{0.10}Zn_{0.90}F_4$. The incoming neutron energy was 4.2 meV. The arrows mark the multimer transitions (D=dimer, T=trimer) with energies indicated by the subscript zJ.

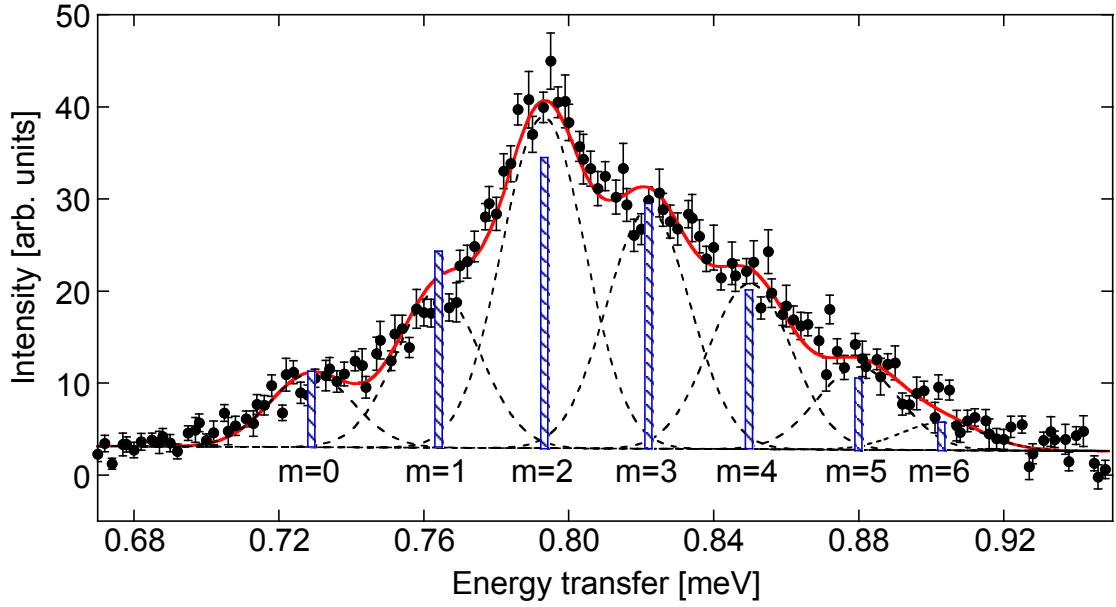

FIG. 2. (Color online) Energy distribution of the $Mn^{2+}$ singlet-triplet dimer transition observed for $KMn_{0.10}Zn_{0.90}F_3$ at T=2 K. The lines are the result of a least-squares fitting procedure as explained in the text. The vertical bars correspond to the probabilities $p_m(x)$ predicted by Eq. (1) with n=26.

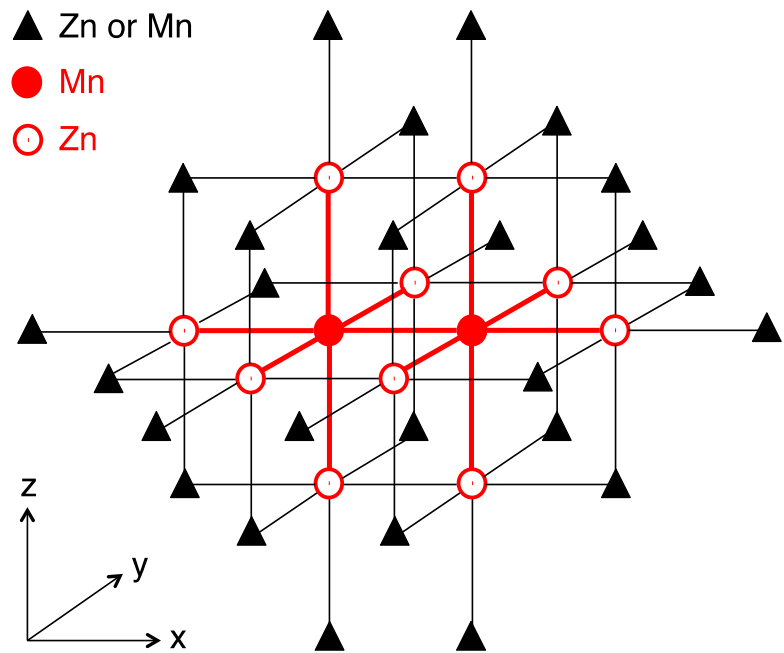

FIG. 3. (Color online) Three-dimensional sketch of the structure of $KMn_xZn_{1-x}F_3$ (only the Mn and Zn positions are shown). The triangles denote the 26 nearest-neighbor positions around the central $Mn^{2+}$ dimer marked by spheres.

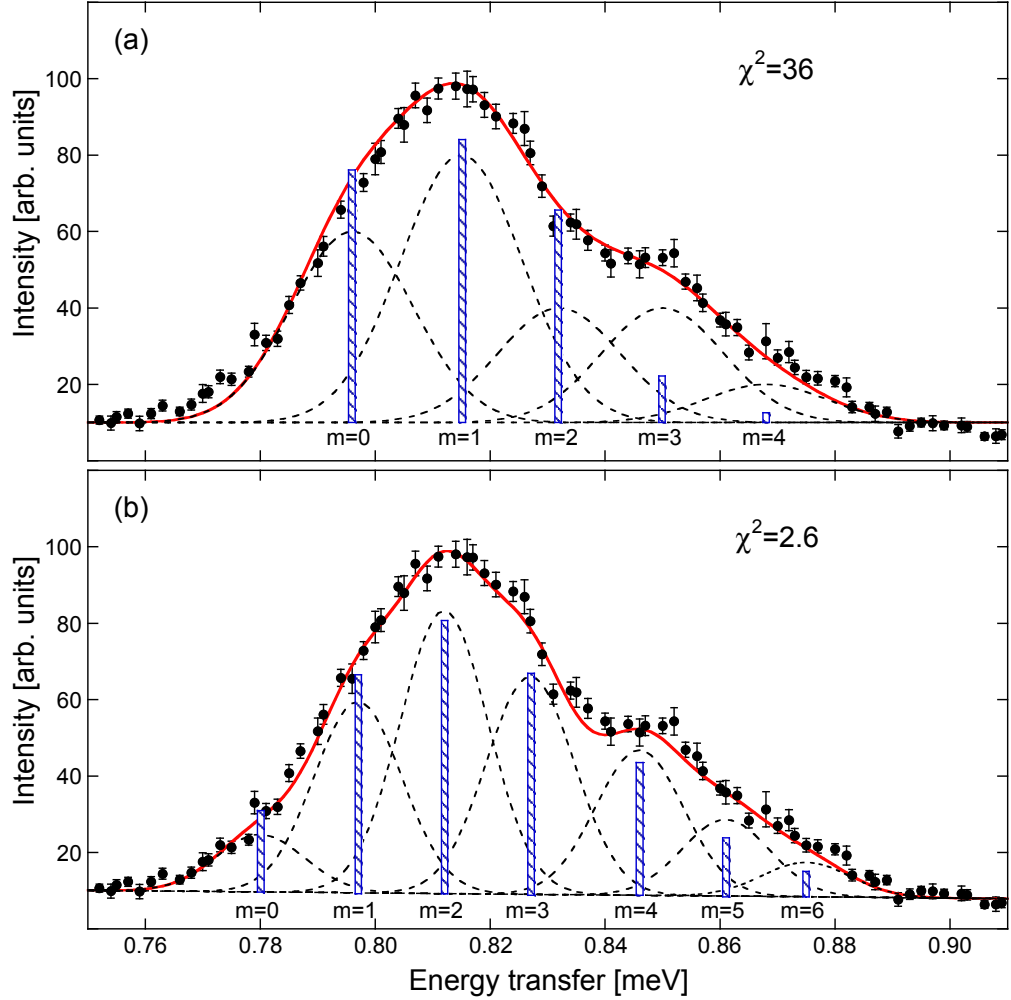

FIG. 4. (Color online) Energy distribution of the $Mn^{2+}$ singlet-triplet dimer transition observed for $K_2Mn_{0.10}Zn_{0.90}F_4$ at T=2 K. The lines are the result of a least-squares fitting procedure as explained in the text. The vertical bars correspond to the probabilities $p_m(x)$ predicted by Eq. (1). The goodness of fit $\chi^2$ is a measure of the agreement between the probabilities $p_m(x)$ and the amplitudes of the individual lines. (a) Statistical model with n=10 nearest-neighbor positions. (b) Statistical model with n=24 nearest-neighbor and next-nearest-neighbor positions.

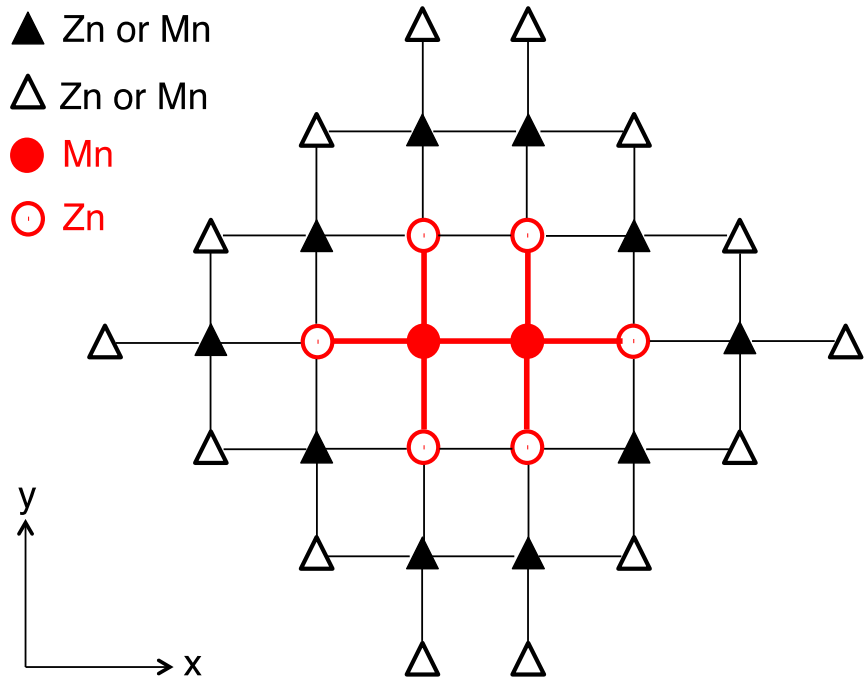

FIG. 5. (Color online) Two-dimensional sketch of the structure of $K_2Mn_xZn_{1-x}F_4$ (only the Mn and Zn positions are shown). The full and open triangles denote the 10 nearest-neighbor and 14 next-nearest-neighbor positions, respectively, around the central $Mn^{2+}$ dimer marked by spheres.